\documentclass[twocolumn,showpacs,preprintnumbers,amsmath,amssymb]{revtex4}

\usepackage{graphicx}
\usepackage{dcolumn}
\usepackage{bm}

\begin{document}


\title{Phase mixing of shear Alfv\'en waves as a new mechanism for electron acceleration
in collisionless, kinetic plasmas}

\author{David Tsiklauri$^1$, Jun-Ichi Sakai$^2$ and Shinji Saito$^2$}
\affiliation{$^1$Institute for Materials Research,
School of Computing, Science and Engineering,
University of Salford, Gt Manchester, M5 4WT, United Kingdom.\\
$^2$Laboratory for Plasma Astrophysics, Faculty of Engineering, Toyama University,
3190, Gofuku, Toyama, 930-8555,  Japan}%
\date{\today}

\begin{abstract}
Particle-In-Cell (kinetic) simulations of shear Alfv\'en wave interaction with 
one dimensional, across the uniform magnetic field,  
density inhomogeneity (phase mixing) in collisionless plasma were performed for the first time.
As a result a new electron acceleration mechanism is discovered.
Progressive distortion of Alfv\'en wave front, due to the differences in 
local Alfv\'en speed, generates nearly parallel to the magnetic
field electrostatic fields, which accelerate electrons via Landau damping.
Surprisingly, amplitude decay law in the inhomogeneous regions, 
in the kinetic regime, is the same as in the MHD approximation 
described by Heyvaerts and Priest (1983).

\end{abstract}


\maketitle

Interaction of shear Alfv\'en waves (AWs) 
with plasma inhomogeneities is a topic
of considerable importance both in astrophysical and laboratory plasmas.
This is due to the fact both AWs and inhomogeneities 
often coexist in many of these physical systems.
shear AWs are believed to be
good candidates for plasma heating, energy and momentum transport.
On the one hand, in many physical situations AWs are easily excitable
and they are present in a number of astrophysical systems.
On the other hand, these waves dissipate on shear viscosity as opposed to
compressive fast and slow magnetosonic waves which dissipate on bulk viscosity.
In astrophysical plasmas shear viscosity is extremely small
as compared to bulk one. Hence,
AWs are notoriously difficult to dissipate.
One of the possibilities to improve AW dissipation is to introduce progressively
collapsing spatial scales, $\delta l \to 0$, into the system (recall that the 
classical viscous and ohmic dissipation is $\propto \delta l^{-2}$). 
Heyvaerts and Priest have proposed (in astrophysical context) one such mechanism called 
AW phase mixing \cite{hp83}. It occurs when a linearly polarised shear
AW propagates in the plasma with one dimensional, transverse to the uniform magnetic 
field density inhomogeneity.
In such situation initially plane AW front is progressively 
distorted because of different Alfv\'en speeds across the field. 
This 
creates progressively strong gradients
across the field (effectively in the inhomogeneous regions transverse scale collapses),
and thus in the case of finite resistivity, dissipation is greatly enhanced.
Thus, it is believed that phase mixing can provide substantial plasma
heating. 
A significant amount of work has been done in the context of heating open
magnetic structures in the solar 
corona \cite{hp83,nph86,p91,nrm97,dmha00,bank00,tan01,hbw02,tn02,tna02,tnr03}.
All phase mixing studies so far have been performed in the MHD approximation,
however, since the transverse scales in the AW collapse progressively to zero,
MHD approximation is inevitably violated, first, when the transverse scale approaches
ion gyro-radius $r_i$ and then electron gyro-radius $r_e$.
Thus, we proposed to study phase mixing effect in the kinetic regime, i.e.
we go beyond MHD approximation.
As a result we discovered new mechanism of  electron acceleration
due to wave-particle interactions which has important implications
for various space and laboratory plasmas, e.g. coronal heating problem and acceleration of solar wind.

We used 2D3V, the fully relativistic, electromagnetic, particle-in-cell (PIC)
code with MPI parallelisation, modified from 3D3V TRISTAN code \cite{b93}.
The system size is $L_x=5000 \Delta$ and $L_y=200 \Delta$, where
$\Delta(=1.0)$ is the grid size. The periodic boundary conditions for
$x$- and $y$-directions are imposed on particles and fields. There are about
478 million electrons and ions in the simulation. The average number of
particles per cell is 100 in low density regions (see below). Thermal velocity of electrons is $v_{th,e}=0.1c$
and for ions is $v_{th,i}=0.025c$.
The ion to electron
mass ratio is $m_i/m_e=16$. The time step is $\omega_{pe} \Delta t=0.05$. Here
$\omega_{pe}$ is the electron plasma frequency.
The Debye length is $v_{th,e}/\omega_{pe}=1.0$. The electron skin depth 
is $c/\omega_{pe}=10 \Delta$, while the ion skin depth is $c/\omega_{pi}=40 \Delta$.
Here $\omega_{pi}$ is the ion plasma frequency.
The electron Larmor radius is $v_{th,e}/\omega_{ce}=1.0 \Delta$, while
the same for ions is $v_{th,i}/\omega_{ci}=4.0 \Delta$.
The external uniform magnetic field, $B_0$,
is in the $x$-direction and the initial
electric field is zero. 
The ratio of electron cyclotron frequency to electron plasma
frequency is $\omega_{ce}/\omega_{pe}=1.0$, while the same for ions is
$\omega_{ci}/\omega_{pi}=0.25$. The latter ratio is essentially $V_A/c$ -- the Alfv\'en
speed. Plasma $\beta=2(\omega_{pe}/\omega_{ce})^2(v_{th,e}/c)^2=0.02$.
Here all plasma parameters are quoted far away from the density 
inhomogeneity region. The dimensionless (normalised to some reference constant value of $n_0$) 
ion and electron density inhomogeneity is described by
\begin{equation}
 {n_i(y)}=
{n_e(y)}=1+3 \exp\left[-\left({(y-100 \Delta)}/{(50 \Delta)}\right)^6\right]
\equiv F(y).
\end{equation}
This means that in the central region (across $y$-direction) density is
smoothly enhanced by a factor of 4, and there are strong density gradients of 
width of about ${50 \Delta}$ around the points $y=51.5 \Delta$ and $y=148.5 \Delta$.
The background temperature of ions and electrons, and their thermal velocities
are varied accordingly
\begin{equation}
{T_i(y)}/{T_0}=
{T_e(y)}/{T_0}=F(y)^{-1},
\end{equation}
\begin{equation}
 {v_{th,i}}/{v_{i0}}=
{v_{th,e}}/{v_{e0}}=F(y)^{-1/2},
\end{equation}
such that the thermal pressure remains constant. Since the background magnetic field
along $x$-coordinate  is also constant, the total pressure remains constant too.
Then we impose current of the following form
\begin{equation}
{\partial_t E_y}=-J_0\sin(\omega_d t)\left(1-\exp\left[-(t/t_0)^2\right]\right),
\end{equation}
\begin{equation}
{\partial_t E_z}=-J_0\cos(\omega_d t)\left(1-\exp\left[-(t/t_0)^2\right]\right).
\end{equation}
Here $\omega_d$ is the driving frequency which was fixed at $\omega_d=0.3\omega_{ci}$,
which ensures that no significant ion-cyclotron damping is present. Also,
$\partial_t$ denotes time derivative.
$t_0$ is the onset time of the driver, which was fixed at $50 /\omega_{pe}$
or $3.125 / \omega_{ci}$. This means that the driver onset time is about 3 ion-cyclotron
periods. Imposing such current on the system results in the generation of
left circularly polarised shear AW, which is driven at the left boundary of simulation box and has a width of $1 \Delta$.
Initial amplitude of the current is such that relative AW amplitude is about 5 \% of the background in the low density regions,
thus the simulation is weakly non-linear.

Because of no initial (perpendicular to the external magnetic field) velocity excitation
was imposed in addition to the above specified currents (cf. \cite{tn02}), 
the excited (driven) at the left boundary circularly polarised AW
is split into two circularly polarised AWs that travel in opposite directions. 
The dynamics of these
waves is shown in Fig.1 where we show three snapshots of the evolution.
Typical simulation, till the last snapshot shown in the figure, takes about 8 days
on the parallel 32 dual 2.4 GHz Xeon  processors.
It can be seen from the figure that because of the periodic boundary conditions circularly polarised
AW that was travelling to the left has reappeared from the right side of the simulation box 
($t=15.62 / \omega_{ci}$).
Then the dynamics of the AW ($B_z,E_y$) progresses in a similar manner as in MHD, i.e. it phase mixes \cite{hp83}.
In other words middle region (in $y$-coordinate) travels slower because of the density enhancement (note that 
$V_A(y) \propto 1/\sqrt{n_i(y)}$).
This obviously causes distortion of initially plane wave front and the creation of strong gradients
in the regions around $y=50$ and $150$.
In MHD approximation, in the case of finite resistivity $\eta$, in these regions
AW is strongly dissipated due to enhanced viscous and ohmic dissipation. 
This effectively means that the outer and inner parts of the
travelling AW are detached from each other and propagate independently.
This is why the effect is called phase mixing -- after long time, in the case
of developed phase mixing, phases in the wave front become effectively uncorrelated.
{\it A priori} it was not clear what to expect from our PIC simulation
because it was performed for the first time. The code is collisionless and there
are no sources of dissipation in it (apart from possibility of wave-particle interactions).
It is evident from Fig.1 that at later stages 
($t=54.69 / \omega_{ci}$) AW front is strongly damped in the strong density
gradient regions.
This immediately rises a question of where AW energy went?
The answer lies in Fig. 2, where we plot $E_x$ longitudinal
electrostatic field, and electron phase space ($V_x/c$ vs. $x$) for the different 
times (note, that in order to reduce figure size, only electrons with $V_x > 0.15c$
were plotted).
In the 
regions around $y=50$ and $150$ for later times significant electrostatic field
is generated. This is the consequence of stretching of AW front in those regions
because of difference in local Alfv\'en speed.
In the right column of this figure we see that exactly in those regions
where $E_x$ is generated, the electrons are accelerated in large numbers.
Thus, we conclude that energy of the phase mixed AW goes into acceleration of electrons.
In fact line plots of $E_x$ show that this electrostatic field is strongly damped.
i.e. energy is channelled to electrons via Landau damping.

The next piece of evidence comes from looking at the distribution function of electrons
before and after the phase mixing took place.
In Fig.3 we plot distribution function of electrons at $t=0$ and $t=54.69 / \omega_{ci}$.
Note that even at $t=0$  the distribution function does not look as purely Maxwellian because
of the fact that temperature varies across $y-$coordinate (to keep total pressure
constant) and the graph is produced for the entire simulation domain.
There is also a substantial difference at $t=54.69 / \omega_{ci}$
to its original form because of the aforementioned electron acceleration.
We see that the number of electrons having velocities $V_x=\pm (0.1-0.3)c$ is increased.
Note that the acceleration of electrons takes place mostly along
the external magnetic field (along $x$-coordinate). No electron acceleration 
occurs in $V_y$ or $V_z$ (not plotted here).

The next step is to check whether the increase in electron velocities comes from the
resonant wave particle interactions. For this purpose in Fig. 4 we plot two snapshots of
Alfv\'en wave $B_z(x,y=148)$ components at instances $t=54.69 / \omega_{ci}$ (solid line)
and $t=46.87 / \omega_{ci}$ (dotted line).
The distance between the two upper leftmost peaks (which is the distance travelled by the
wave in time between the snapshots) 
is about $\delta L=150\Delta=15(c/\omega_{pe})$.
Time difference between the snapshots is $\delta t=7.82 / \omega_{ci}$.
Thus, measured AW speed at the point of the strongest density gradient ($y=148$)
is $V_A^M=\delta L /\delta t=0.12c$. Now we can also work out 
the Alfv\'en speed.
In the homogeneous low density region the Alfv\'en speed was set to be
$V_A(\infty)=0.25 c$. From Eq.(1) it follows that for $y=148$ density is increased by a factor of
$2.37$ which means that the Alfv\'en speed at this position is
$V_A(148)=0.25/\sqrt{2.37}c=0.16c$.
The measured and calculated Alfv\'en speeds in the inhomogeneous region do not coincide. 
This can be attributed to the fact that in the inhomogeneous regions
(where electron acceleration takes place) because of momentum conservation
AW front is decelerated as it passes on energy and momentum to the
electrons. However, this may be not the case if wave-particle interactions
play the same role as dissipation in the MHD:
Then wave-particle interactions would result only in the decrease of the AW
amplitude (dissipation) not in its deceleration.
If we compare these values to Fig.3, we deduce that these are the
velocities $>0.12c$ above which electron numbers with higher velocities
are greatly increased. This deviation peaks at about $0.25c$ which
in fact corresponds to the Alfv\'en speed in the lower density regions.
This can be explained by the fact the electron acceleration takes
place in  wide regions (cf. Fig. 2) along and around $y=148$ (and $y=51$) -- thus
the spread in the accelerated velocities.

In Fig.4 we also plot a visual fit curve (dashed line) in order to 
quantify the amplitude decay law for the AW (at $t=54.69 / \omega_{ci}$)
in the strongest density inhomogeneity region.
The fitted (dashed) cure is represented by $0.056 \exp \left[ -
\left({x}/{1250}\right)^3\right]$.
There is an astonishing similarity of this fit with the 
MHD approximation results.
Authors of Ref. \cite{hp83} found that for large times (developed phase mixing),
in the case of harmonic driver, the amplitude decay law
is given by $\propto \exp \left[ -
\left(\frac{\eta \omega^2 V_A^{\prime 2}}{6 V_A^{5}}\right)x^3\right]$ which is much faster 
than the usual resistivity dissipation
$\propto \exp(-\eta x)$. Here $V_A^{\prime}$ is the derivative
of the Alfv\'en speed with respect to $y$-coordinate.
The most intriguing fact is that even in the kinetic approximation
the same $\propto \exp (-A x^3)$ law holds as in the MHD.
In the MHD a finite resistivity and
Alfv\'en speed non-uniformity are responsible for the
enhanced dissipation via phase mixing mechanism.
In our PIC simulations (kinetic phase mixing), however, we do not have dissipation
and collisions (dissipation). Thus, in our case
wave-particle interactions play the same role as 
resistivity $\eta$ in the MHD phase mixing.
It should be noted that no significant AW dissipation
was found away from the density inhomogeneity regions.
This has the same explanation as in the case of MHD --
it is the regions of density of inhomogeneities ($V_A^{\prime}\not=0$) where
the dissipation is greatly enhanced, while in the regions
where $V_A^{\prime}=0$ there is no substantial dissipation (apart from classical $\propto \exp(-\eta x)$
one).
In MHD approximation aforementioned amplitude decay law is derived
from the diffusion equation, to which MHD equations reduce to for large times (developed
phase mixing). It seems that kinetic description 
leads to the same type of diffusion equation.
It is unclear although, at this stage, what physical quantity 
would play role of resistivity $\eta$ (from the MHD approximation) in the
kinetic regime. 

It is worthwhile to mention that in MHD approximation 
authors of  Refs. \cite{hbw02,tnr03} showed that 
in the case of localised Alfv\'en pulses,
Heyvaerts and Priest's amplitude decay
formula $\propto \exp (-A x^3)$ (which is true for
harmonic AWs) is replaced by the power law $B_z \propto x^{-3/2}$. 
A natural next step forward
would be to check whether 
in the case of localised Alfv\'en pulses the same power law holds
in the kinetic regime.

Finally we would like to mention that after this study was complete
we became aware of study by \citet{vh04}, who used hybrid
code (electrons treated as neutralising fluid, while ion
kinetics is retained) as opposed to our (fully kinetic) PIC code,
to simulate resonant absorption. They found that 
a planar (body) Alfv\'en
wave propagating at less than $90^{\circ}$ to a background gradient  
has field lines which lose wave energy to another set of field lines by
cross-field transport. Further, \citet{v04} found that 
when perpendicular scales of
order 10 proton inertial lengths ($10 c/\omega_{pi}$) develop from wave refraction
in the vicinity of the resonant field lines, a non-propagating density fluctuation begins
to grow to large amplitudes. This saturates by exciting highly oblique, compressive, and
low-frequency waves which dissipate and heat protons. 
These processes lead to a faster development of small
scales across the magnetic field, i.e. this is ultimately related to the 
phase mixing mechanism, studied here.
Continuing this argument we would like make clear distinction
between the effects of phase mixing and resonant absorption of the shear Alfv\'en waves.
Historically, there was some confusion in use of there terms. In fact, 
earlier studies which were performed in the context of laboratory plasmas, e.g.
Ref. \cite{hc74} who proposed (quote) "the heating of collisionless plasma by 
utilising a spatial phase mixing by shear Alfv\'en wave resonance and discussed potential 
applications to toroidal plasma" used term phase mixing to discuss the physical
effect which in reality was the resonant absorption. In their later works, \citet{hc75} and 
\citet{hc76}, when they treated the same problem in the kinetic approximation 
(note that \citet{hc74} used MHD approach) 
the authors avoided further use of term phase mixing and instead they talked about
linear mode conversion of shear Alfv\'en wave into kinetic Alfv\'en wave near the resonance
$\omega^2=k_{\parallel}^2V_A(x)$ (in their geometry the inhomogeneity was across the $x$-axis).
In the kinetic approximation \citet{hc75} and 
\citet{hc76} found a number of interesting effects including:

(i) They established that near the resonance initial shear Alfv\'en wave
is linearly converted into kinetic Alfv\'en wave (which is the Alfv\'en wave modified by
the finite ion Larmor radius and electron inertia) that propagates almost
parallel to the magnetic field into the higher density side. 
Inclusion of the finite ion Larmor radius and electron inertia removes the logarithmic singularity
present in the MHD resonant absorption which exists because in MHD
shear Alfv\'en wave cannot propagate across the magnetic field.
Physically this means that finite ion Larmor radius
prevents ting ions to magnetic field lines and thus allows
propagation across the field. The electron inertia eliminates singularity
in the same fashion but on different, $r_i\sqrt{T_e/T_i}$, spatial scale 
(here $r_i$ ion gyro-radius). 

(ii) They also found that in collionless regime (which is applicable to our case)
dissipation of the (mode converted) kinetic Alfv\'en wave is through
Landau damping of parallel electric fields both by electrons and ions.
However, in the low plasma $\beta$ regime (which is also applicable to our case
$\beta=2(\omega_{pe}/\omega_{ce})^2(v_{th,e}/c)^2=0.02$) they showed that
only electrons are heated (accelerated), but not the ions.
In our numerical simulations we see similar behaviour: i.e. preferential
acceleration of electrons (cf. \citet{tss05}). In spite of this similarity, however, one
should make clear distinction between (a) phase mixing which essentially
is a result of enhanced, due to plasma inhomogeneity, {\it collisional}
viscous and ohmic dissipation and (b) resonant absorption which can operate
in the {\it collisionless} regime as a result of generation of kinetic
Alfv\'en wave in the resonant layer, which subsequently decays via Landau
damping preferentially accelerating electrons in the low $\beta$ regime.

(iii) They also showed that in the kinetic regime the total absorption rate
is approximately the same as in the MHD. Our simulations seem to produce
similar results. We cannot comment quantitatively on this occasion,
but at least the spatial from of the amplitude decay ($\propto \exp (-A x^3)$) is similar
in both cases.

Resuming aforesaid, we conjecture that the generated nearly parallel
electrostatic fields found in our numerical simulations are due to the
generation of kinetic Alfv\'en waves that are created through interaction
of initial shear Alfv\'en waves with plasma density inhomogeneity,
in a similar fashion as in resonant
absorption described above. Further theoretical study is thus needed to
provide solid theoretical basis for interpretation of 
our numerical simulation results and to test the above conjecture.

\begin{acknowledgments}
The authors would like to express their special gratitude to 
CAMPUS (Campaign to Promote University of Salford) which funded
J.-I.S.'s one month fellowship to the Salford University 
that made this project possible.
DT acknowledges use of E. Copson Math cluster 
funded by PPARC and University of St. Andrews.
DT kindly acknowledges support from Nuffield Foundation 
through an award to newly appointed lecturers in science,
engineering and mathematics (NUF-NAL 04).
DT would like to thank A.W. Hood (St. Andrews) for encouraging discussions.
We also would like to thank the referees, especially referee 1, for very
useful suggestions.

\end{acknowledgments}

\bibliography{xxx}

\begin{thebibliography}{18}
\expandafter\ifx\csname natexlab\endcsname\relax\def\natexlab#1{#1}\fi
\expandafter\ifx\csname bibnamefont\endcsname\relax
  \def\bibnamefont#1{#1}\fi
\expandafter\ifx\csname bibfnamefont\endcsname\relax
  \def\bibfnamefont#1{#1}\fi
\expandafter\ifx\csname citenamefont\endcsname\relax
  \def\citenamefont#1{#1}\fi
\expandafter\ifx\csname url\endcsname\relax
  \def\url#1{\texttt{#1}}\fi
\expandafter\ifx\csname urlprefix\endcsname\relax\def\urlprefix{URL }\fi
\providecommand{\bibinfo}[2]{#2}
\providecommand{\eprint}[2][]{\url{#2}}

\bibitem[{\citenamefont{Heyvaerts and Priest}(1983)}]{hp83}
\bibinfo{author}{\bibfnamefont{J.}~\bibnamefont{Heyvaerts}} \bibnamefont{and}
  \bibinfo{author}{\bibfnamefont{E.}~\bibnamefont{Priest}},
  \bibinfo{journal}{Astron. \ Astrophys.} \textbf{\bibinfo{volume}{117}},
  \bibinfo{pages}{220} (\bibinfo{year}{1983}).

\bibitem[{\citenamefont{Nocera et~al.}(1986)\citenamefont{Nocera, Priest, and
  Hollweg}}]{nph86}
\bibinfo{author}{\bibfnamefont{L.}~\bibnamefont{Nocera}},
  \bibinfo{author}{\bibfnamefont{E.}~\bibnamefont{Priest}}, \bibnamefont{and}
  \bibinfo{author}{\bibfnamefont{J.}~\bibnamefont{Hollweg}},
  \bibinfo{journal}{Geophys. \ Astrophys. \ Fl. Dyn.}
  \textbf{\bibinfo{volume}{35}}, \bibinfo{pages}{111} (\bibinfo{year}{1986}).

\bibitem[{\citenamefont{Parker}(1991)}]{p91}
\bibinfo{author}{\bibfnamefont{E.}~\bibnamefont{Parker}},
  \bibinfo{journal}{Astrophys. \ J.} \textbf{\bibinfo{volume}{376}},
  \bibinfo{pages}{355} (\bibinfo{year}{1991}).

\bibitem[{\citenamefont{Nakariakov et~al.}(1997)\citenamefont{Nakariakov,
  Roberts, and Murawski}}]{nrm97}
\bibinfo{author}{\bibfnamefont{V.}~\bibnamefont{Nakariakov}},
  \bibinfo{author}{\bibfnamefont{B.}~\bibnamefont{Roberts}}, \bibnamefont{and}
  \bibinfo{author}{\bibfnamefont{K.}~\bibnamefont{Murawski}},
  \bibinfo{journal}{Sol. \ Phys.} \textbf{\bibinfo{volume}{175}},
  \bibinfo{pages}{93} (\bibinfo{year}{1997}).

\bibitem[{\citenamefont{Moortel et~al.}(2000)\citenamefont{Moortel, Hood, and
  Arber}}]{dmha00}
\bibinfo{author}{\bibfnamefont{I.~D.} \bibnamefont{Moortel}},
  \bibinfo{author}{\bibfnamefont{A.~W.} \bibnamefont{Hood}}, \bibnamefont{and}
  \bibinfo{author}{\bibfnamefont{T.}~\bibnamefont{Arber}},
  \bibinfo{journal}{Astron. \ Astrophys.} \textbf{\bibinfo{volume}{354}},
  \bibinfo{pages}{334} (\bibinfo{year}{2000}).

\bibitem[{\citenamefont{Botha et~al.}(2000)\citenamefont{Botha, Arber,
  Nakariakov, and Keenan}}]{bank00}
\bibinfo{author}{\bibfnamefont{G.~J.~J.} \bibnamefont{Botha}},
  \bibinfo{author}{\bibfnamefont{T.~D.} \bibnamefont{Arber}},
  \bibinfo{author}{\bibfnamefont{V.}~\bibnamefont{Nakariakov}},
  \bibnamefont{and} \bibinfo{author}{\bibfnamefont{F.}~\bibnamefont{Keenan}},
  \bibinfo{journal}{Astron. \ Astrophys.} \textbf{\bibinfo{volume}{363}},
  \bibinfo{pages}{1186} (\bibinfo{year}{2000}).

\bibitem[{\citenamefont{Tsiklauri et~al.}(2001)\citenamefont{Tsiklauri, Arber,
  and Nakariakov}}]{tan01}
\bibinfo{author}{\bibfnamefont{D.}~\bibnamefont{Tsiklauri}},
  \bibinfo{author}{\bibfnamefont{T.}~\bibnamefont{Arber}}, \bibnamefont{and}
  \bibinfo{author}{\bibfnamefont{V.~M.} \bibnamefont{Nakariakov}},
  \bibinfo{journal}{Astron. \ Astrophys.} \textbf{\bibinfo{volume}{379}},
  \bibinfo{pages}{1098} (\bibinfo{year}{2001}).

\bibitem[{\citenamefont{Hood et~al.}(2002)\citenamefont{Hood, Brooks, and
  Wright}}]{hbw02}
\bibinfo{author}{\bibfnamefont{A.}~\bibnamefont{Hood}},
  \bibinfo{author}{\bibfnamefont{S.}~\bibnamefont{Brooks}}, \bibnamefont{and}
  \bibinfo{author}{\bibfnamefont{A.~N.} \bibnamefont{Wright}},
  \bibinfo{journal}{Proc. \ Roy. \ Soc. \ Lond. \ A}
  \textbf{\bibinfo{volume}{458}}, \bibinfo{pages}{2307} (\bibinfo{year}{2002}).

\bibitem[{\citenamefont{Tsiklauri and Nakariakov}(2002)}]{tn02}
\bibinfo{author}{\bibfnamefont{D.}~\bibnamefont{Tsiklauri}} \bibnamefont{and}
  \bibinfo{author}{\bibfnamefont{V.~M.} \bibnamefont{Nakariakov}},
  \bibinfo{journal}{Astron. \ Astrophys.} \textbf{\bibinfo{volume}{393}},
  \bibinfo{pages}{321} (\bibinfo{year}{2002}).

\bibitem[{\citenamefont{Tsiklauri et~al.}(2002)\citenamefont{Tsiklauri,
  Nakariakov, and Arber}}]{tna02}
\bibinfo{author}{\bibfnamefont{D.}~\bibnamefont{Tsiklauri}},
  \bibinfo{author}{\bibfnamefont{V.~M.} \bibnamefont{Nakariakov}},
  \bibnamefont{and} \bibinfo{author}{\bibfnamefont{T.}~\bibnamefont{Arber}},
  \bibinfo{journal}{Astron. \ Astrophys.} \textbf{\bibinfo{volume}{395}},
  \bibinfo{pages}{285} (\bibinfo{year}{2002}).

\bibitem[{\citenamefont{Tsiklauri et~al.}(2003)\citenamefont{Tsiklauri,
  Nakariakov, and Rowlands}}]{tnr03}
\bibinfo{author}{\bibfnamefont{D.}~\bibnamefont{Tsiklauri}},
  \bibinfo{author}{\bibfnamefont{V.~M.} \bibnamefont{Nakariakov}},
  \bibnamefont{and} \bibinfo{author}{\bibfnamefont{G.}~\bibnamefont{Rowlands}},
  \bibinfo{journal}{Astron. \ Astrophys.} \textbf{\bibinfo{volume}{400}},
  \bibinfo{pages}{1051} (\bibinfo{year}{2003}).

\bibitem[{\citenamefont{Buneman}(1993, p.67)}]{b93}
\bibinfo{author}{\bibfnamefont{O.}~\bibnamefont{Buneman}},
  \emph{\bibinfo{title}{in Computer Space Plasma Physics: Simulation Techniques
  and Software}} (\bibinfo{publisher}{Terra Scientific}, \bibinfo{address}{New
  York}, \bibinfo{year}{1993, p.67}).

\bibitem[{\citenamefont{Vasquez and Hollweg}(2004)}]{vh04}
\bibinfo{author}{\bibfnamefont{B.~J.} \bibnamefont{Vasquez}} \bibnamefont{and}
  \bibinfo{author}{\bibfnamefont{J.~V.} \bibnamefont{Hollweg}},
  \bibinfo{journal}{Geophys. \ Res. \ Lett.} \textbf{\bibinfo{volume}{31}},
  \bibinfo{pages}{L14803} (\bibinfo{year}{2004}).

\bibitem[{\citenamefont{Vasquez}(2004)}]{v04}
\bibinfo{author}{\bibfnamefont{B.~J.} \bibnamefont{Vasquez}},
  \bibinfo{journal}{J.\ Geophys. \ Res. (submitted)}  (\bibinfo{year}{2004}).

\bibitem[{\citenamefont{Hasegawa and Chen}(1974)}]{hc74}
\bibinfo{author}{\bibfnamefont{A.}~\bibnamefont{Hasegawa}} \bibnamefont{and}
  \bibinfo{author}{\bibfnamefont{L.}~\bibnamefont{Chen}},
  \bibinfo{journal}{Phys. \ Rev. \ Lett.} \textbf{\bibinfo{volume}{32}},
  \bibinfo{pages}{454} (\bibinfo{year}{1974}).

\bibitem[{\citenamefont{Hasegawa and Chen}(1975)}]{hc75}
\bibinfo{author}{\bibfnamefont{A.}~\bibnamefont{Hasegawa}} \bibnamefont{and}
  \bibinfo{author}{\bibfnamefont{L.}~\bibnamefont{Chen}},
  \bibinfo{journal}{Phys. \ Rev. \ Lett.} \textbf{\bibinfo{volume}{35}},
  \bibinfo{pages}{370} (\bibinfo{year}{1975}).

\bibitem[{\citenamefont{Hasegawa and Chen}(1976)}]{hc76}
\bibinfo{author}{\bibfnamefont{A.}~\bibnamefont{Hasegawa}} \bibnamefont{and}
  \bibinfo{author}{\bibfnamefont{L.}~\bibnamefont{Chen}},
  \bibinfo{journal}{Phys. \ Fluids.} \textbf{\bibinfo{volume}{19}},
  \bibinfo{pages}{1924} (\bibinfo{year}{1976}).

\bibitem[{\citenamefont{Tsiklauri et~al.}(2005)\citenamefont{Tsiklauri, Sakai,
  and Saito}}]{tss05}
\bibinfo{author}{\bibfnamefont{D.}~\bibnamefont{Tsiklauri}},
  \bibinfo{author}{\bibfnamefont{J.~I.} \bibnamefont{Sakai}}, \bibnamefont{and}
  \bibinfo{author}{\bibfnamefont{S.}~\bibnamefont{Saito}},
  \bibinfo{journal}{Astron. \ Astrophys. \ (accepted)}  (\bibinfo{year}{2005}).

\end{thebibliography}

\begin{figure*}
\centering
 \includegraphics[width=12cm]{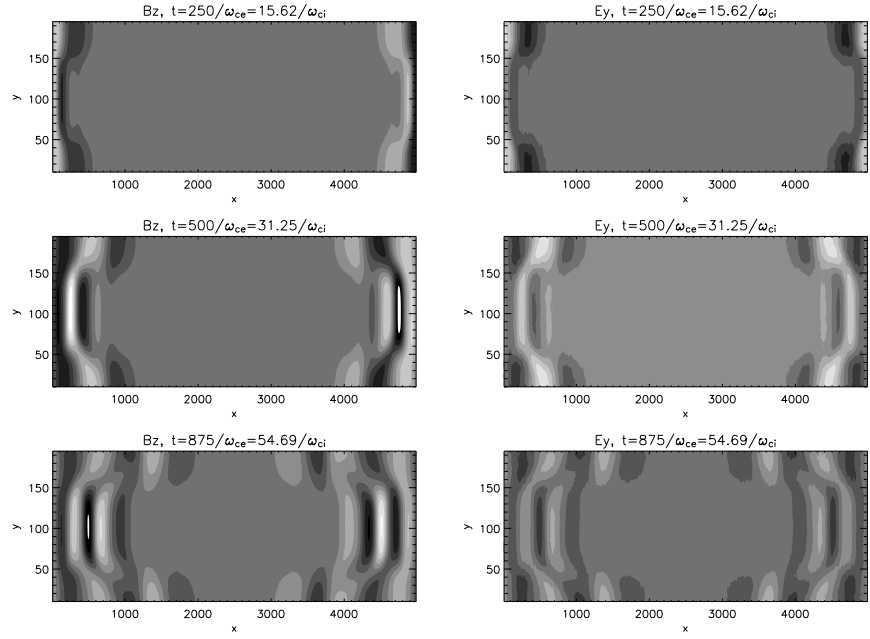}
\caption{Contour (intensity) plots of phase mixed Alfv\'en wave $B_z$ and $E_y$ components
at instants: $t=(15.62, 31.25, 54.69) / \omega_{ci}$. Excitation source is at the left
boundary. Because of periodic boundary conditions, left-propagating AW re-appears from the 
right side of the simulation box. Note how AW is progressively stretched because of 
differences in local Alfv\'en speed.}
\end{figure*}

\begin{figure*}
\centering
\includegraphics[width=12cm]{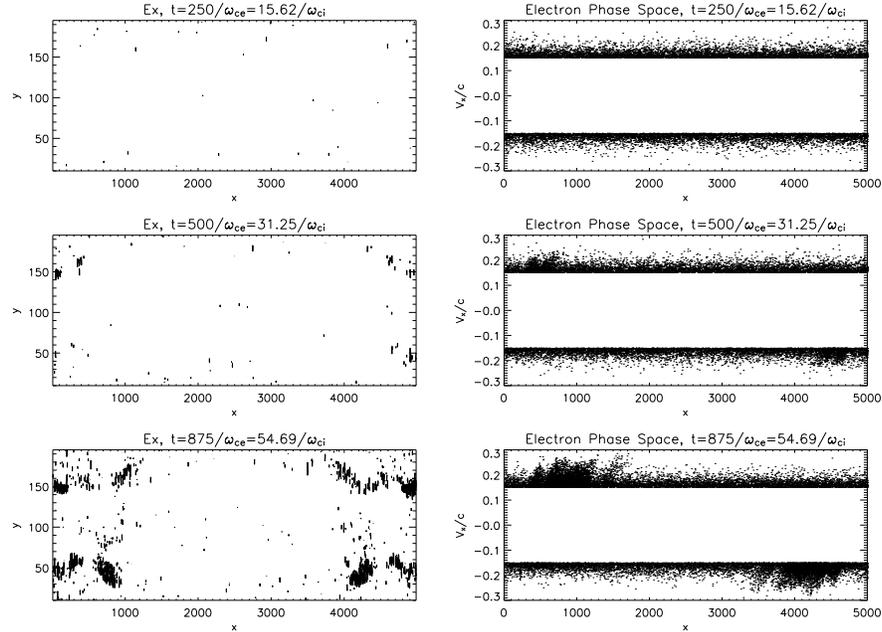}
\caption{Left column: contour plots of generated nearly parallel to the
external magnetic field electrostatic field $E_x$ at instants: 
$t=(15.62, 31.25, 54.69) / \omega_{ci}$. Right column: $x$-component of electron phase
space at the same times. Note, that in order to reduce figure size, only electrons with $V_x > 0.15c$
were plotted.}
\end{figure*}

\begin{figure}[]
\resizebox{\hsize}{!}{\includegraphics{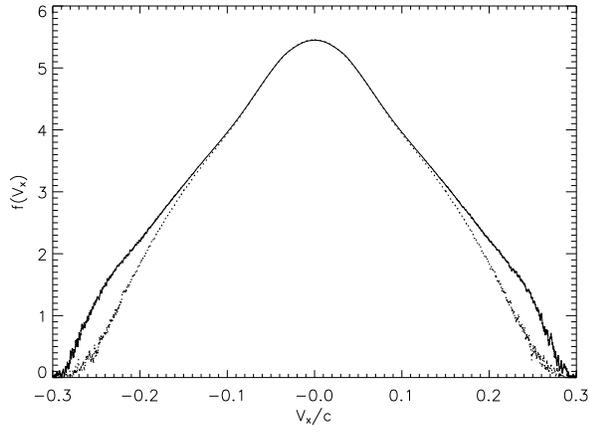}} 
\caption{The distribution function of electrons at $t=0$ (dotted curve) and $t=54.69 / \omega_{ci}$
(solid curve).}
\end{figure}

\begin{figure}[]
\resizebox{\hsize}{!}{\includegraphics{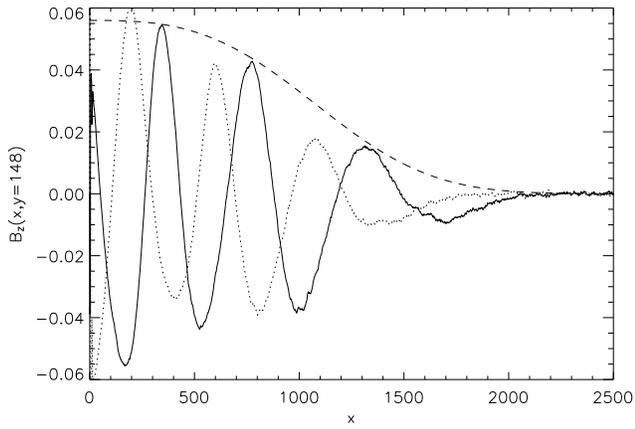}} 
\caption{Two snapshots of
Alfv\'en wave $B_z(x,y=148)$ component at instants $t=54.69 / \omega_{ci}$ (solid line)
and $t=46.87 / \omega_{ci}$ (dotted line). Dashed line represents fit
$0.056 \exp \left[ -\left({x}/{1250}\right)^3\right]$.}
\end{figure}
\end{document}